\documentclass[twocolumn,prb,amssymb,showpacs]{revtex4}
\usepackage{graphicx}

\begin{document}

\title{Quantum corrections to conductivity: from weak to
strong localization}

\author{G.~M.~Minkov}
\email{Grigori.Minkov@usu.ru}
\author{O.~E.~Rut}
\author{A.~V.~Germanenko}
\author{A.~A.~Sherstobitov}
\affiliation{Institute of Physics and Applied Mathematics, Ural
State University, 620083 Ekaterinburg, Russia}

\author{B.~N.~Zvonkov}
\author{E.~A.~Uskova}
\author{A.~A.~Birukov}
\address{Physical-Technical Research Institute, University of
Nizhni Novgorod, 603600 Nizhni Novgorod, Russia}

\date{\today}

\begin{abstract}
Results of detailed investigations of the conductivity and Hall effect
in gated single quantum well GaAs/InGaAs/GaAs heterostructures with
two-dimensional electron gas are presented. A successive analysis of
the data has shown that the conductivity is diffusive for $k_F l=25-2$.
The absolute value of the quantum corrections for $k_F l = 2$ at low
temperature is not small, e.g., it is about $70$~\% of the Drude
conductivity at $T=0.46$~K. For $k_F l=2-0.5$ the conductivity looks
like diffusive one. The temperature and magnetic field dependences are
qualitatively described within the framework of the self-consistent
theory by Vollhardt and W\"{o}lfle. The interference correction is
therewith close in magnitude to the Drude conductivity so that the
conductivity $\sigma$ becomes significantly less than $e^{2}/h$. We
conclude that the temperature and magnetic field dependences of
conductivity in the whole $k_Fl$ range are due to changes of quantum
corrections.
\end{abstract}

 \maketitle

\section{Introduction}
\label{sec:int}

The weak localization regime in two-dimensional (2D) systems at $k_F
l\gg 1$ ($k_F$ and $l$ are the Fermi quasimomentum and mean free path,
respectively), when the electron motion is diffusive, is well
understood from theoretical point of view.\cite{Altshuler} In this case
the quantum corrections to conductivity, which are caused by
electron-electron interaction and interference, are small compared with
the Drude conductivity $\sigma_0=\pi G_0 k_F l $, where
$G_0=e^2/(2\pi^2\hbar)$. Experimentally, this regime was studied in
different types of 2D systems. Qualitative and in some cases
quantitative agreement with the theoretical predictions was found.
However, important questions remain to be answered: (i) what does
happen with these corrections at decrease of $k_F l$ down to $k_F
l\simeq 1$ and (ii) at decrease of temperature $T$ when the quantum
corrections  become comparable in magnitude\footnote{It should be noted
that the term ``corrections" is not quite adequate in this case.
Nevertheless, we will use it even though the correction magnitude is
not small.} with the Drude conductivity?

It is clear that the increase of disorder sooner or later leads to the
change of the conductivity mechanism from diffusive one to hopping and
another question is: when does it happen? To answer this question the
temperature dependence of the conductivity is analyzed usually. It  is
supposed that the transition to the hopping conductivity occurs when
the conductivity becomes lower than $e^2/h$ and the strong temperature
dependence arises.\cite{h1,h2,h3,h4,h5} From our point of view, it is
not enough to analyze only the $\sigma(T)$ dependence. The low
conductivity and its strong temperature dependence can result from
large value of the quantum corrections as well.

The aim of this paper is to study the role of quantum corrections over
wide range of $k_F l$ and to understand what happens when the quantum
corrections become comparable with the Drude conductivity. We try to
answer these questions studying the quantum corrections in the 2D
systems with the simplest and well-known electron energy spectrum,
starting from the well understandable case $k_F l\gg 1$.

We report experimental results obtained for gated single quantum well
GaAs/InGaAs/GaAs structures with one 2D subband occupied. An analysis
of the experimental data  shows that the conductivity is diffusive when
$k_F l$ varies from $25$ to approximately $2$ and looks like diffusive
one at $k_Fl=2-0.5$. At low temperature and $k_F l\simeq 1$ the total
quantum correction is not small. It is close in magnitude to the Drude
conductivity so that the conductivity is significantly less than
$e^{2}/h\simeq 3.86\times 10^{-5}$~$\Omega^{-1}$ at low temperature:
for instance, for $k_F l\simeq 0.5$ it is about $3\times
10^{-8}$~$\Omega^{-1}$ at $T=0.46$~K, that is six hundred odd times
less than $\sigma_0$. Thus, in this range the strong temperature and
magnetic field dependences are caused by the change of quantum
corrections.

\section{samples}
The heterostructures investigated were grown by metal-organic
vapor-phase epitaxy on a semiinsulator GaAs substrate.  They
consist of $0.5$-mkm-thick undoped GaAs epilayer, a Sn $\delta$
layer, a 60-\AA\ spacer of undoped GaAs, a 80-\AA\
In$_{0.2}$Ga$_{0.8}$As well, a 60-\AA\ spacer  of undoped GaAs, a
Sn $\delta$ layer, and a 3000-\AA\ cap layer of undoped GaAs. The
samples were mesa etched into standard Hall bars with the
dimensions $1.2\times 0.2$ mm$^2$ and then the gate electrode was
deposited onto the cap layer by thermal evaporation. The
measurements were performed in the temperature range $0.4-12$ K at
magnetic field $B$ up to 6 T. The electron density was found from
the Hall effect and from the Shubnikov-de Haas oscillations where
it was possible. These values coincide with an accuracy of 5\%.

\begin{figure}
\includegraphics[width=\linewidth,clip=true]{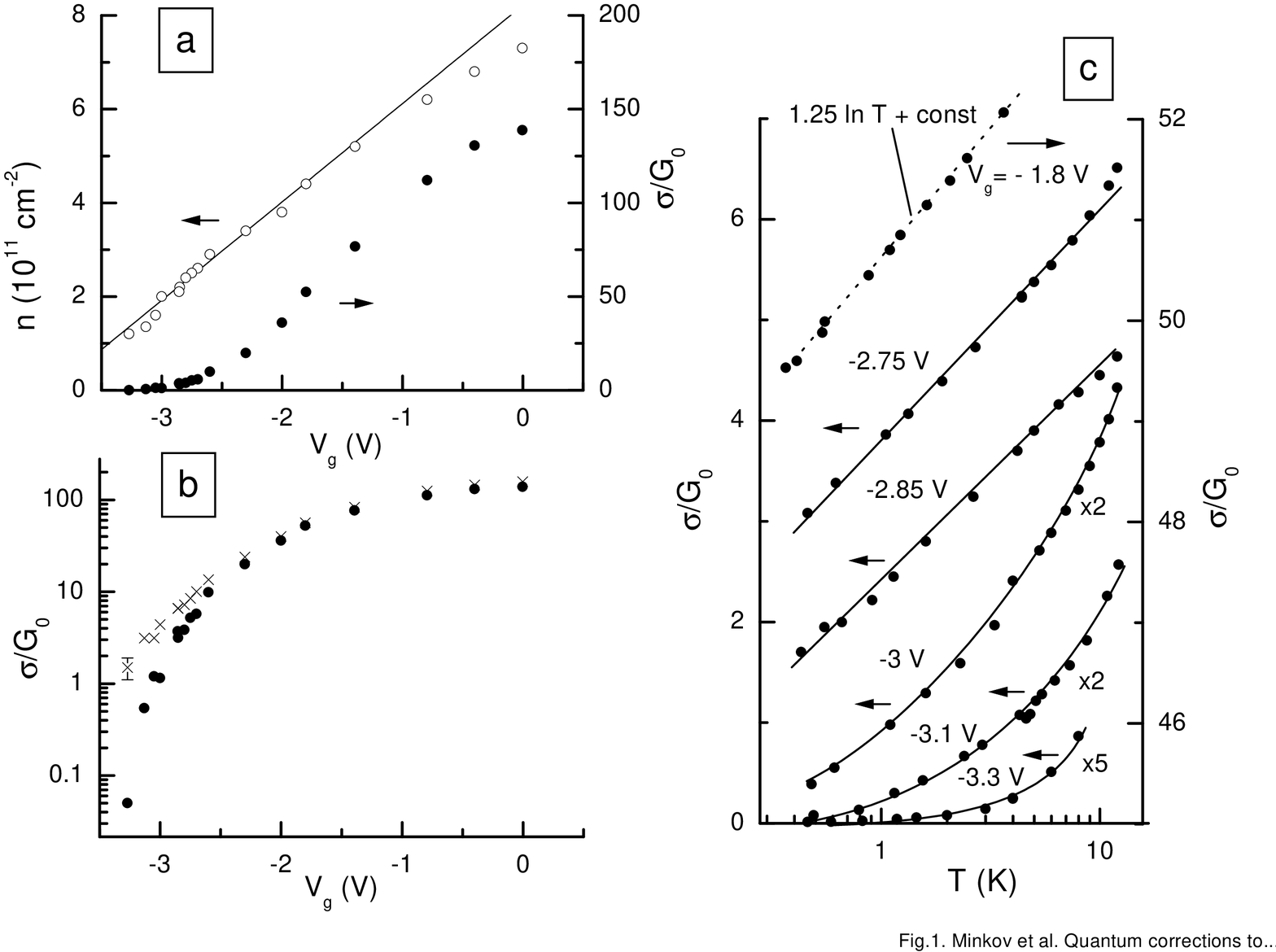}
\caption{(a) The gate voltage dependence of the conductivity at
$T=4.2$~K (solid circles) and electron density in quantum well (open
circles). The line is the dependence $n_t(V_g)$ calculated from the
electrostatic consideration (see text). (b) The gate voltage
dependences of conductivity at $T=4.2$~K (circles) and the Drude
conductivity (crosses). (c) Temperature dependences of the conductivity
for different gate voltages. The solid lines are the guide for an eye,
dotted line is Eq.~(\ref{eq07}) with $(1+3/4\lambda)=0.35$ and $p=0.9$
(see text).} \label{fig1}
\end{figure}

The gate voltage dependences of the electron density and conductivity
are presented  in Fig.\ \ref{fig1}(a),~(b). Varying the gate voltage
$V_g$ from $0.0$ to $-3.3$ V we changed the electron density in the
quantum well from $7.5\times 10^{11}$ to $1.2\times 10^{11}$ cm$^{-2}$
and the conductivity $\sigma$ at $T= 4.2$~K from $2.1\times 10^{-3}$ to
$6\times 10^{-7}$ $\Omega^{-1}$. The straight line in Fig.\
\ref{fig1}(a) shows the $V_g$-dependence of the total electron density
in the quantum well and $\delta$ layers, calculated from the simple
electrostatic consideration: $n_t(V_g)=n(0)+V_g\, C/|e|$ with $n(0)$ as
fitting parameter. Here, $C$ is the gate-2D channel capacity per
centimeter squared $C=\varepsilon/(4\pi d)$, where $d=3000$\AA\ is the
cap-layer thickness and $\varepsilon=12.5$. The deviation of the
experimental data from the line, which is evident  at $V_g>-1$~V is a
result of that the fraction of electrons occupies the states in
$\delta$ layers. Arising of the electrons and, hence, empty states at
the Fermi energy in $\delta$ layers leads to specific features in
transport, \cite{Tau-phi} that need an additional detailed
investigation and will be discussed elsewhere.

\section{Results and discussion}
The temperature dependences of the conductivity for some gate voltages
are presented in Fig.\ \ref{fig1}(c). It is clearly seen that for
$V_g\geq -2.85$ V  the temperature dependences of $\sigma$ are close to
the logarithmic ones. For lower $V_g$, when the conductivity is less
than $e^2/h$, the significant deviation from logarithm is evident. The
conductivity in this range is usually interpreted as the hopping
conductivity. \cite{h1,h2,h3,h4,h5} Below we will show that the quantum
corrections can lead to such a behavior if they become close in
magnitude to the Drude conductivity.

To clarify the role of quantum corrections at low conductivity
when $k_F l \simeq 1$ let us analyze the experimental data
starting from $k_F l \gg 1$ when the conventional theories of the
quantum corrections are applicable. Following the sequence of data
treatment  described in Ref.\ \onlinecite{e-e} we will assure at
first that the quantum correction theories describe temperature,
low and high magnetic field behavior of the conductivity. After
that we will find the contributions of the electron-electron
interaction and quantum interference to the conductivity and then
trace their changes with lowering of $k_F l$.

\subsection{The case $k_F l \gg 1$}
 \label{subsec1}
\begin{figure}
\includegraphics[width=\linewidth,clip=true]{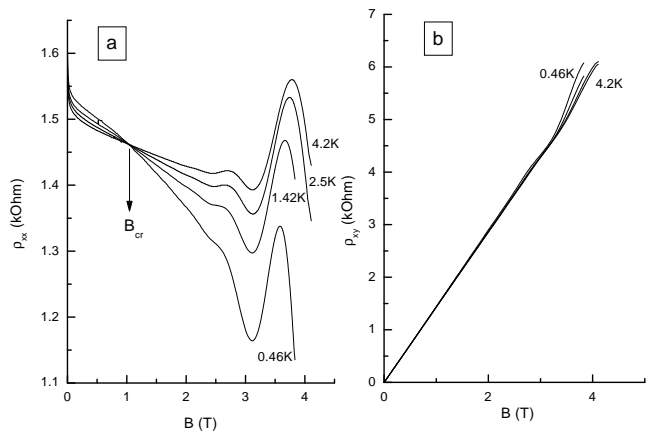}
\caption{The magnetic field dependences of $\rho_{xx}$ (a) and
$\rho_{xy}$ (b) for different temperatures at $V_g=-1.8$ V ($k_F
l=17.9$).} \label{fig2}
\end{figure}

In Fig.\ \ref{fig2} the magnetic field dependences of $\rho_{xx}$ and
$\rho_{xy}$ taken at different temperatures for gate voltage $V_g=-1.8$
V are presented. The two different magnetic field ranges are evident in
Fig.\ \ref{fig2}(a): the range of sharp decrease of $\rho_{xx}$ at low
field $B\leq 0.1-0.2$ T, and the range of moderate dependence at higher
field. All the $\rho_{xx}$-versus-$B$ curves cross each other at fixed
magnetic field $B_{cr}$.

The high-magnetic field behavior of $\rho_{xx}$ fully corresponds
to the theory for electron-electron interaction\cite{Altshuler}
which predicts that
\begin{eqnarray}
\delta\sigma_{xx}^{ee}(T)&=&G_0
\left(1+\frac{3}{4}\lambda\right)\ln\left(\frac{
 kT\tau}{\hbar}\right) \label{eq00} \\ \delta\sigma_{xy}^{ee}&=&0
\label{eq01}
\end{eqnarray}
and hence
\begin{equation}
\rho_{xx}(B,T)\simeq
\frac{1}{\sigma_0}-\frac{1}{\sigma_0^2}\left(1-\mu^2
B^2\right)\delta\sigma_{xx}^{ee}(T), \label{eq02}
\end{equation}
when $\delta\sigma_{xx}^{ee}\ll \sigma_0$. Here, $\tau$ is the
quasimomentum relaxation time, and $\lambda$ is the parameter of
electron-electron interaction. The value of $\lambda$ has been
calculated in Ref.\ \onlinecite{Fin} and found being independent of the
magnetic field when $g \mu_B B\leq kT$. It is clearly seen from
Eq.~(\ref{eq02}) that $\rho_{xx}$-versus-$B$ plots measured for
different temperatures have to cross each other at magnetic field
$B=\mu^{-1}$. Experimentally, the value of $B_{cr}=1$ T is really close
to $1/\mu$ with $\mu=0.99$ m$^2/(\text{V}\times \text{sec})$.

\begin{figure}
\includegraphics[width=\linewidth,clip=true]{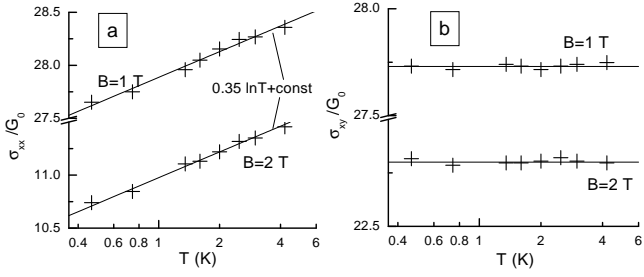}
\caption{Temperature dependence of $\protect\sigma_{xx}$ (a) and $\protect%
\sigma_{xy}$ (b) for two magnetic fields, $V_g=-1.8$~V} \label{fig3}
\end{figure}

The temperature dependences of $\sigma_{xx}$ and $\sigma_{xy}$ for high
magnetic field are presented in Fig.\ \ref{fig3}. As seen $\sigma_{xy}$
is temperature independent within an experimental error. The
temperature dependence of $\sigma_{xx}$ is close to the logarithmic
one. The slope of the $\sigma_{xx}$-versus-$\ln\, T$ plot does not
depend on the magnetic field. Thus, the high magnetic field behavior of
conductivity tensor components  agrees completely with the theoretical
predictions for the correction due to electron-electron interaction. It
allows us to determine the value of $(1+3/4\lambda)$ [see Eq.\
(\ref{eq00})]. So, for $V_g=-1.8$~V we obtain
$(1+3/4\lambda)=0.35\pm0.05$.
\begin{figure}
\includegraphics[width=\linewidth,clip=true]{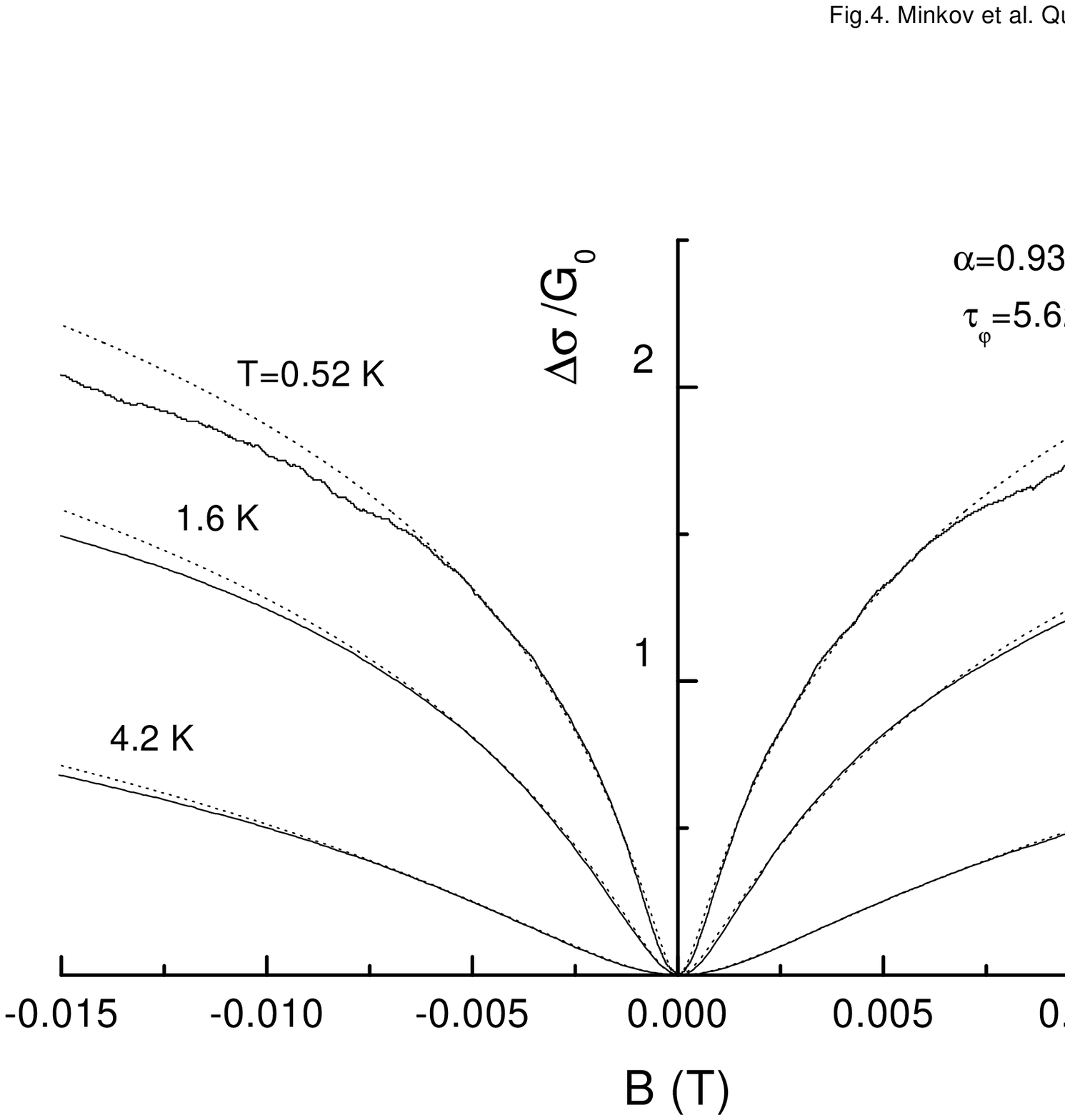}
\caption{The magnetic field dependence of
$\Delta\sigma(B)=1/\rho_{xx}(B)-1/\rho_{xx}(0)$ for different
temperatures, $V_g=-1.8$~V. Solid curves are the experimental data.
Dashed lines are the best fit to Eq. (\ref{eq03}) made over the
magnetic field range from 0 to $0.25\,B_{tr}$, $B_{tr}=0.03$~T. The
parameters of the best fit are given near the corresponding curves. }
\label{fig4}
\end{figure}

We turn now to the low-magnetic-field behavior of $\rho_{xx}$, which is
a consequence of suppression of the interference correction by magnetic
field (Fig.\ \ref{fig4}). At $B <0.5B_{tr}$, where $B_{tr}=\hbar
/(2el^2)$,\cite{ftn1} the dependences
$\Delta\sigma(B)=1/\rho_{xx}(B)-1/\rho_{xx}(0)$ are described by the
well-known expression\cite{Hik}
\begin{eqnarray}
\Delta\sigma(B)&=&\alpha G_0 \biggl\{ \psi\left(\frac{1}{2}+\frac{%
\tau}{\tau_\varphi}\frac{B_{tr}}{B}\right)\nonumber \\
&- &\psi\left(\frac{1}{2}+\frac{%
B_{tr}}{B}\right)- \ln{\left(\frac{\tau}{\tau_\varphi}\right)}
\biggr\} \label{eq03}
\end{eqnarray}
with $\alpha$ and $\tau_\varphi$ given in this figure. In
Eq.~(\ref{eq03}), $\psi(x)$ is a digamma function, $\tau_\varphi$ is
the phase breaking time. A difference of the prefactor $\alpha$ from
unity, which is more pronounced at higher temperature, is sequence of
low ratio $\tau_\varphi/\tau$. For instance, $\tau_\varphi/\tau\simeq
20$ for $T=4.2$~K. It is apparently not enough for the diffusion
approximation.\cite{our1} Nevertheless, as shown in Ref.\
\onlinecite{our1} the use of Eq.~(\ref{eq03}) for the fit of
experimental data in this regime gives the value of $\tau_\varphi$ very
close to the true one.

\begin{figure}
\includegraphics[width=6cm,clip=true]{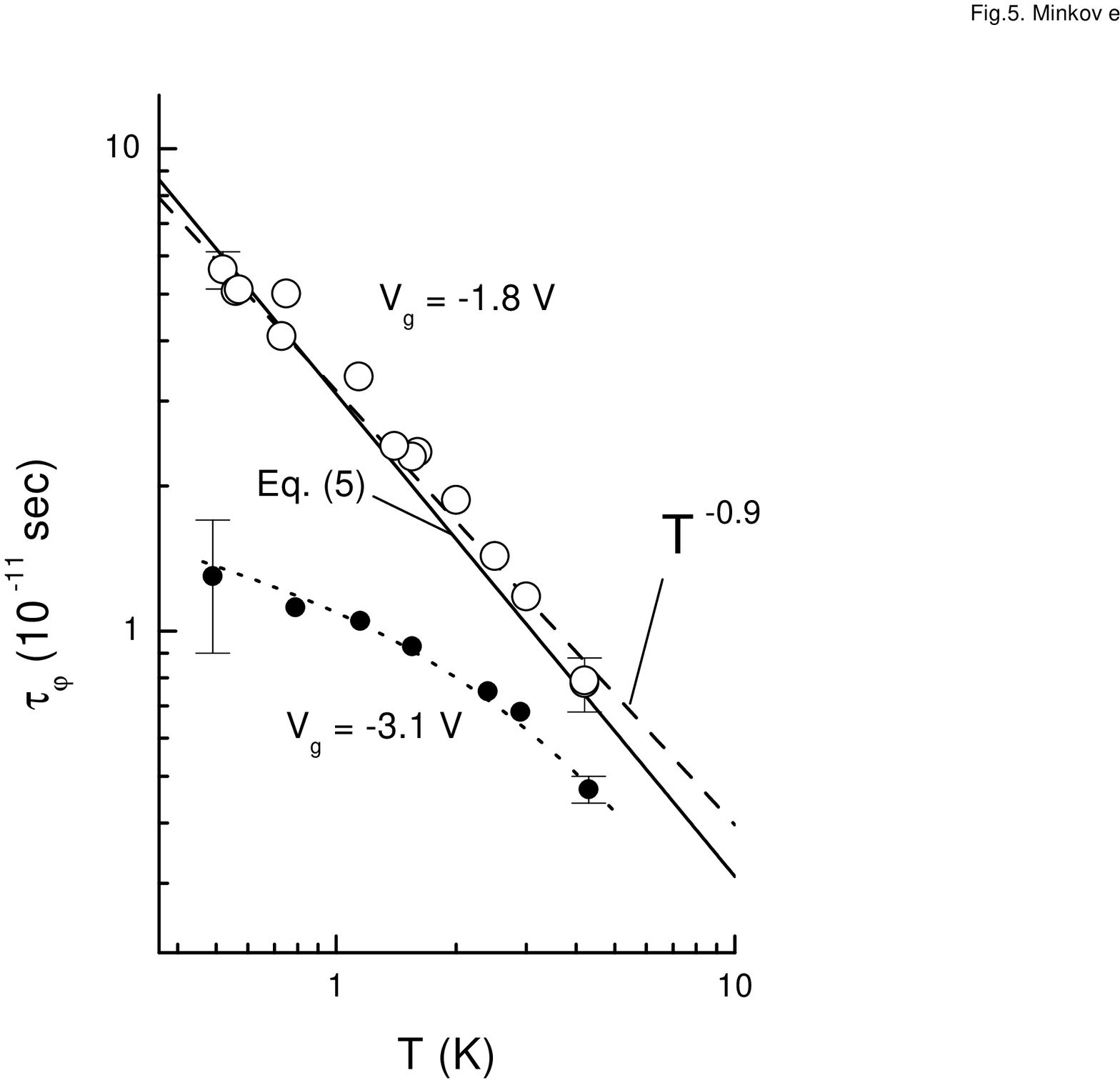}
\caption{The temperature dependences of $\tau_\varphi$ at $V_g=-1.8$~V,
$k_F l=17.9$ (open circles) and $V_g=-3.1$ V, $k_F l\simeq 1 $ (solid
circles). Solid line is the theoretical dependence given by
Eq.~(\ref{eq06}), dashed line is the power law $T^{-0.9}$, dotted line
is the guide for an eye.} \label{fig5}
\end{figure}

If one plots the temperature dependence of $\tau_\varphi$ found
from the fit, we obtain it being close to $T^{-p}$ with $p\simeq
0.9$ (Fig.~\ref{fig5}), which is close to the theoretical value
$p=1$:
\begin{equation}
\tau_\varphi^{-1}= \frac{kT}{\hbar}\frac{2\pi
G_0}{\sigma_0}\ln\left(\frac{ \sigma_0}{2\pi G_0}\right).
\label{eq04}
\end{equation}
Notice, that not only the temperature dependence but the absolute
value of $\tau_\varphi$ obtained experimentally is close to the
theoretical one.

\begin{figure}
\includegraphics[width=\linewidth,clip=true]{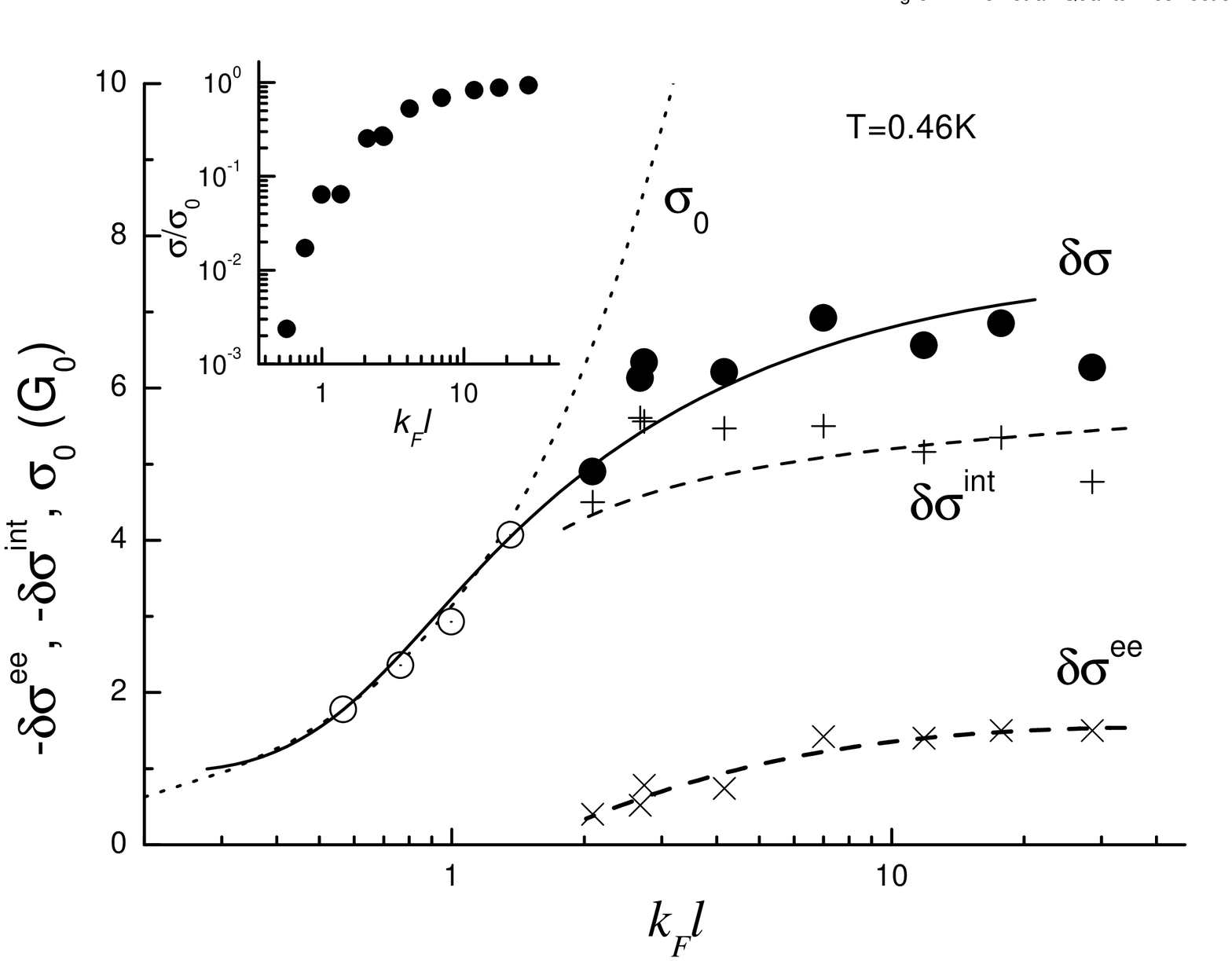}
\caption{The contributions to the conductivity due to electron-electron
interaction $\delta\sigma^{ee}$, interference  $\delta\sigma^{int}$,
and the total contribution $\delta\sigma$ as a function of $k_Fl$. The
dotted curve is $\sigma_0=\pi k_F l$, other curves are the guide for an
eye. Solid and open circles are obtained as described in
Sec.~\ref{subsec1} and Sec.~\ref{subsec2}, respectively. Inset shows
the ratio of the conductivity in zero magnetic field at $T=0.46$~K to
the Drude conductivity plotted as a function of $k_F l$.} \label{fig6}
\end{figure}

So, we can determine the absolute value of the interference correction
at $B=0$ using the well-known expression
\begin{equation}
\delta\sigma^{int}=-G_0 \ln\left(\frac{\tau_\varphi}{\tau}+1
\right). \label{eq05}
\end{equation}
Fig.~\ref{fig6} shows the results obtained for $T=0.46$~K. As seen
$\delta\sigma^{int}$ is practically independent of $k_F l$ while $k_F
l\geq 2$.

Now, when we have experimentally found the values of both
electron-electron and interference contributions to the conductivity,
we can determine the total value of quantum corrections
$\delta\sigma(T)=\delta\sigma^{int}(T)+\delta\sigma^{ee}(T)$ and the
value of the Drude conductivity:
\begin{equation}
 \sigma_0=\sigma(T)- \delta\sigma(T).
 \label{eq06}
\end{equation}
If the experimental results are adequately described by the theory of
the quantum corrections we have to obtain the same values of $\sigma_0$
from the data taken at different temperatures. Really, this procedure
gives the close values. The scatter is about $(0.5-1)\, G_0$ for
$T=0.46 - 10$~K in the whole gate voltage range. The Drude conductivity
obtained by this way as a function of $V_g$ is presented in Fig.\
\ref{fig1}~(b).

The found value of $\sigma_0$ can be further compared with
$\rho_{xx}^{-1}(B_{cr})$. Both quantities have to be equal each other
as it follows from Eq.~(\ref{eq02}). As an example we consider the case
of $V_g=-1.8$ V.  The value of $\sigma_0$ obtained with the help of
Eq.~(\ref{eq06}) is equal to $(56.4\pm 0.5)\, G_0$. Inspection of Fig.\
\ref{fig2} gives $\rho_{xx}(B_{cr})^{-1}=55.6\, G_0$. It is slightly
lower than $\sigma_0$. The reason for this difference is
transparent.\cite{e-e} It is due to the remainder of the interference
correction which is not fully suppressed by magnetic field even at
$B=B_{cr}\simeq 30\, B_{tr}$.

Finally, the temperature dependence of the conductivity at zero
magnetic field is determined by the overall temperature dependences of
$\delta\sigma_{xx}^{ee}$ [Eq.~(\ref{eq00})] and $\delta\sigma^{int}$
[Eq.~(\ref{eq05})]. Thus,
\begin{equation}
\sigma(T)\propto G_0\left(1+\frac{3}{4}\lambda+p\right)\ln\,T.
\label{eq07}
\end{equation}
The dotted line in Fig.\ \ref{fig1}(c) demonstrates a good agreement of
the experimental data with Eq.(\ref{eq07}) when one uses
$(1+3/4\lambda)=0.35$ and $p=0.9$ obtained above.

As is clear from above the cross  of $\rho_{xx}(B)$ for different $T$
is essential for the thorough analysis. Unfortunately the cross point
is observed not over the whole gate voltage range because the
decreasing of $V_{g}$ leads to mobility decrease and hence to shift of
$B_{cr}$ to high magnetic field. At $V_{g}<-2.9$ V the cross is out the
used magnetic field range. But at $V_{g}>-2.9$~V wherever the cross is
evident the good agreement with all the theoretical predictions takes
place. The minimal value of $k_Fl$ therewith is about 2.

Besides, Eq.~(\ref{eq02}) is valid when one can ignore the quantization
of the energy spectrum in a magnetic field, i.e., when the cyclotron
energy $\hbar\omega_c$ is less than the broadening of the Landau levels
or the Fermi energy. In the structure investigated $\hbar\omega_c$ at
$B=\mu^{-1}$ is less than the Fermi energy while $n>2\times
10^{11}$~$\text{cm}^{-2}$ that corresponds to $k_F l>2$.

Thus, we maintain that at $k_Fl\gtrsim 2$ just the quantum corrections
determine the temperature, low and high magnetic field dependences of
the conductivity in two-dimensions. The total value of the corrections
decreases only slightly at decreasing $k_Fl$ (see Fig.~\ref{fig6}), the
contribution due to electron-electron interaction is 25-30~\% of the
interference contribution at $k_Fl\simeq 25$ and only 10~\% at
$k_Fl\simeq 2$. The significant point is that the quantum corrections
at low temperature can be comparable in magnitude with the Drude
conductivity, e.g., at $k_Fl=2.3$ ($V_g=-2.85$ V), their value is about
two-third of $\sigma_0$ when $T=0.46$ K. So, the strong enough
temperature dependence of the conductivity at $B=0$ in this case (see
Fig.~\ref{fig1}~(c)) is caused by the decreasing of the quantum
corrections with temperature.

\begin{figure}
\includegraphics[width=\linewidth,clip=true]{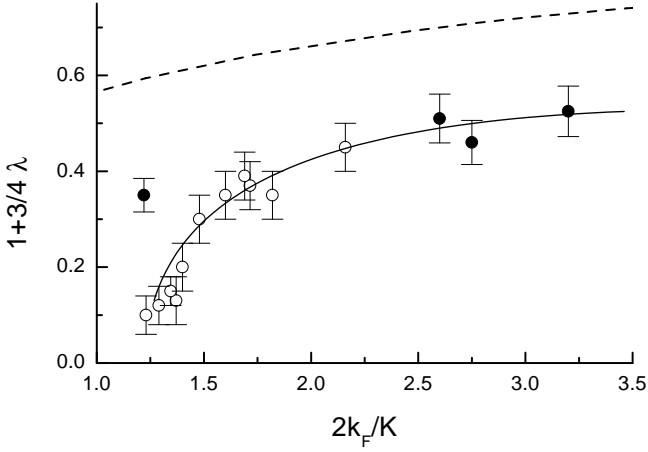}
 \caption{The value of multiplier $\left( 1+3/4\protect\lambda \right)$
in Eq.~(\ref{eq00}) as a function of $2k_{F}/K$. The open circles are
the results obtained for structure investigated, the solid circles are
the results from Ref.\ \protect\onlinecite{e-e}. The dashed curve
represents the theoretical result from Ref.\ \protect\onlinecite{Fin}.
Solid curve is the guide for an eye.} \label{fig7}
\end{figure}

Now we consider the behavior of the electron-electron contribution with
$k_F$ changing. In Fig.~\ref{fig7} the experimental $k_F$-dependences
of $(1+3/4\lambda)$ obtained for structure investigated together with
the theoretical curve\cite{Fin} are presented. One can see that at
$2k_F/K>1.5$ ($K$ is the screening parameter) when  $k_F l>5$ the
experiment lies somewhat below the theoretical curve. Over this range
of $2k_F/K$ our results are close to those from Ref.\ \onlinecite{e-e}.
The strong deviation of  $(1+3/4\lambda)$ at $2k_F/K<1.5$ can be result
of the low value of $k_F l$ whereas the theory has been developed for
$k_F l\gg 1$.

\subsection{The case $k_F l \simeq 1$}
 \label{subsec2}
\begin{figure}
\includegraphics[width=\linewidth,clip=true]{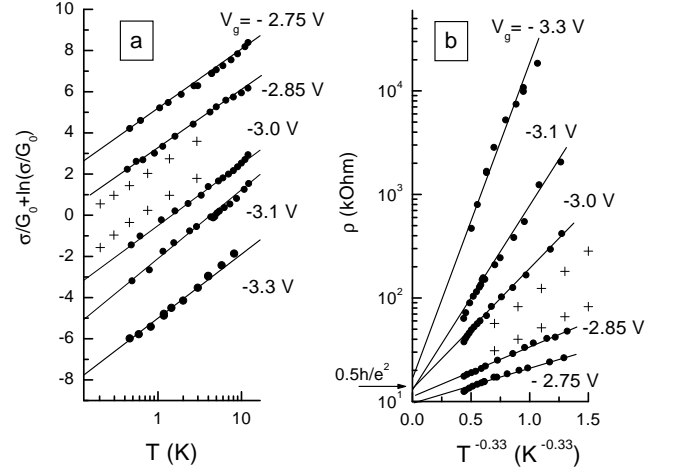}
\caption{The temperature dependence of $\sigma$ (a) and  $\rho$ (b) in
the coordinates corresponding to Eqs.~(\ref{eq09}) and (\ref{eq10})
with $m=1/3$ respectively. The circles are our data, the crosses are
results from Ref.~\protect\onlinecite{h5}. The values of $k_F l$ from
$V_g=-2.75$ V to $-3.3$ V are the following: $k_F l\simeq 3$; $2.1$;
$1.4$; $1.0$; and $0.5$.} \label{fig8}
\end{figure}

Let us analyze the data for $V_g< -2.9$ V when the conductivity is low
and the cross point is not observed. First of all, one can see from
Fig.~\ref{fig1} (c) that the temperature dependence of $\sigma$ is not
logarithmic in this case. It is not surprising because
Eqs.~(\ref{eq00}) and (\ref{eq05}) are valid when the corrections are
small compared with the Drude conductivity. When the temperature tends
to zero, Eq.~(\ref{eq06}) together with Eqs.~(\ref{eq00}) and
(\ref{eq05}) gives the negative value of the conductivity that is
meaningless. It is obvious that another theoretical approach should be
used in this situation. Self-consistent
calculations\cite{Voelfle,Gogolin}  lead to the equation for the
conductivity
\begin{equation}
\frac{\sigma(T)}{\sigma_0}=1-\frac{1}{\pi k_F
l}\ln\left(1+\frac{\tau_\varphi(T)}{2\tau}
\frac{\sigma(T)}{\sigma_0}\right).
 \label{eq08}
\end{equation}
When $\tau_\varphi \gg \tau$ it can be rewritten as
\begin{equation}
\frac{\sigma_0}{G_0}+\ln\left(\frac{\sigma_0}{G_0}\right)-
\left[\frac{\sigma(T)}{G_0}+\ln\left(\frac{\sigma(T)}{G_0}\right)\right]=
\ln\left[\frac{\tau_\varphi(T)}{\tau}\right].
 \label{eq09}
\end{equation}
This equation coincides with Eq.~(\ref{eq05}) if $\sigma_0-\sigma(T)\ll
\sigma_0$, and at $\tau_\varphi(T)/\tau$ going to the infinite gives
$\sigma(T)$ going to zero. In Fig.\ \ref{fig8} we  present our
experimental results as $[\sigma(T)/G_0 + \ln(
\sigma(T)/G_0)]$-versus-$\ln T$ plot in accordance with
Eq.~(\ref{eq09}). It is evident  the experimental data are well
described by this theory.

\begin{figure}
\includegraphics[width=6cm,clip=true]{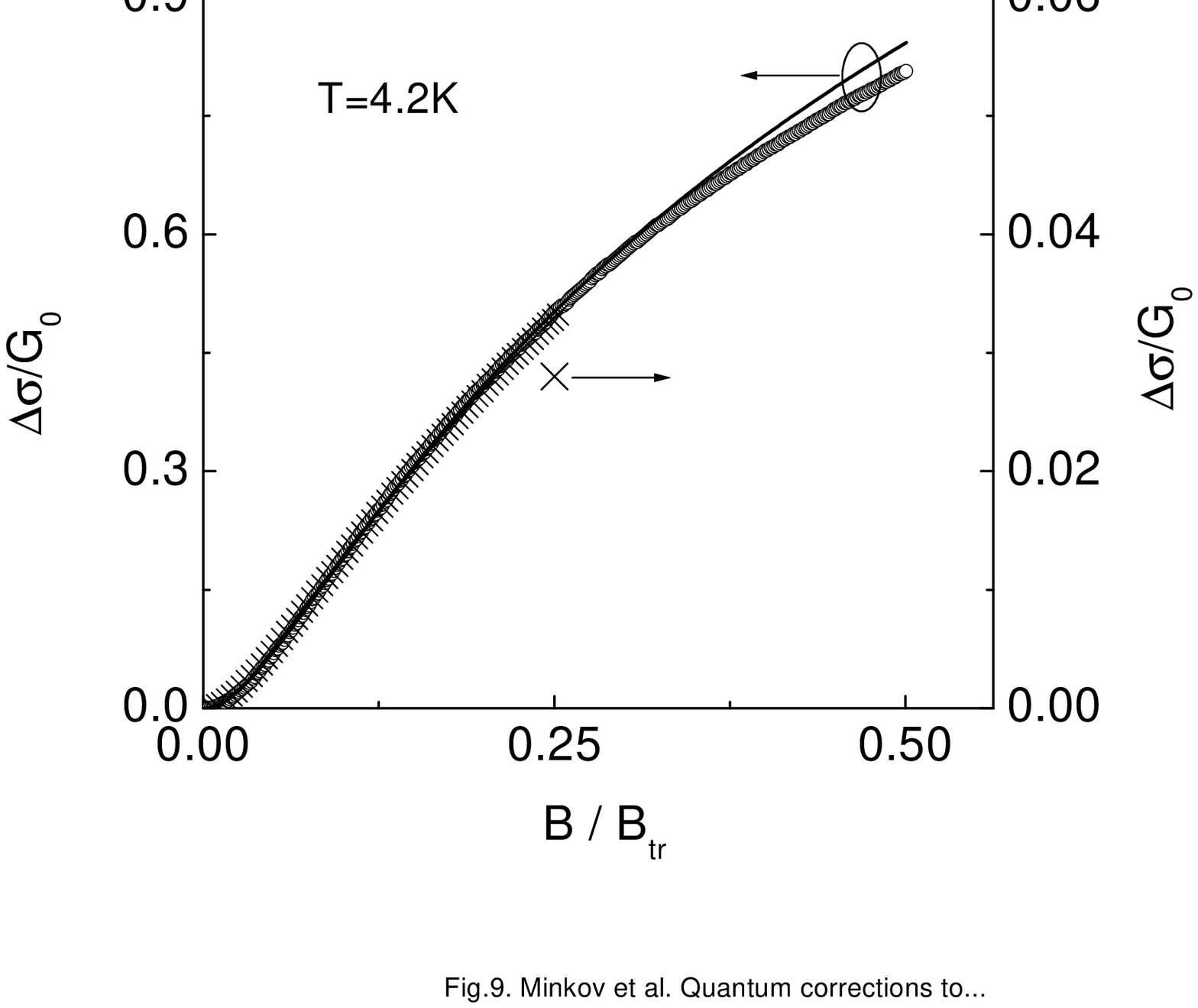}
 \caption{The negative magnetoresistance for $V_g=-1.8$ V
($k_Fl=17.9$) (circles) and $V_g=-3.3$ V ($k_Fl\simeq 0.5$) (crosses),
$T=4.2$~K.} \label{fig9}
\end{figure}

Moreover, the negative magnetoresistance is observed in this range too.
Notice, the shape of $\Delta \sigma$-versus-$B$ dependence is the same
as for large $k_F l$. It is illustrated by Fig.\ \ref{fig9} where
$\Delta\sigma$-versus-$B$ data for $V_g=-1.8$ V ($k_Fl=17.9$) and
$V_g=-3.3$ V ($k_Fl\simeq 0.5$, see below for details) are presented.
It is evident that both data sets practically coincide. From our point
of view this fact indicates that at both small and large $k_Fl$-value
the negative magnetoresistance results from the magnetic-field
suppression of the interference correction to the conductivity.

Let us try to determine the phase breaking time from the negative
magnetoresistance. By analogy with the temperature dependence of
$\sigma$  [see Eq.~(\ref{eq09})] we have analyzed the magnetic filed
dependence of
$\sigma(B)/G_0+\ln[\sigma(B)/G_0]-[\sigma(0)/G_0+\ln[\sigma(0)/G_0]]$
rather than $\sigma(B)-\sigma(0)$ as was in Eq.~(\ref{eq03}). Note,
that the way of finding the value of $B_{tr}$ used in the case $k_Fl\gg
1$ (see Ref.~\onlinecite{ftn1}) is pure now because $\sigma$ strongly
differs from $\sigma_0$. Therefore, we used successive approximation
method. For the first approximation we have used $\sigma(0)$ as
$\sigma_0$, found $B_{tr}$ and, than, determined $\tau_\varphi/\tau$
from the fit of magnetoresistance. After that we have substituted this
ratio into Eq.~(\ref{eq09}) and found the corrected value of $\sigma_0$
and so on. So, output of this procedure is the value of the Drude
conductivity $\sigma_0$ and the ratio $\tau_\varphi/\tau$.

\begin{figure}
\includegraphics[width=\linewidth,clip=true]{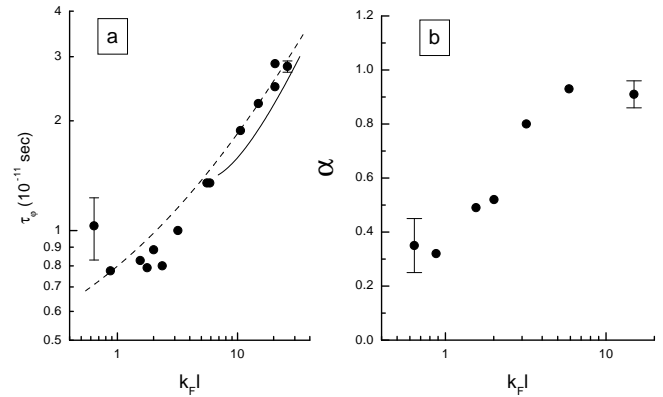}
\caption{The conductivity dependence of $\tau_\varphi$ (a) and
prefactor $\alpha$ (b) at $T=1.5$ K. Symbols are the experimental
results, the solid line in (a) is Eq.~(\ref{eq04}), the dashed line is
the guide for an eye.} \label{fig10}
\end{figure}

We realize that Eqs.~(\ref{eq03}) and (\ref{eq09}) have been used
beyond the framework of their workability. Nevertheless, let us
consider the results. It has been found that the lowering of $V_g$ down
to $-3.3$ V leads to the fall of the Drude conductivity down to $\simeq
1.5\, G_0$ [see Fig.~\ref{fig1}~(b)], that is, $k_F l\simeq 0.5$. The
values of fittimg parameters $\tau_\varphi$ and $\alpha$ therewith
change slowly and monotonically over the whole $k_F l$-range. It is
clearly seen from Fig.~\ref{fig10} in which the results for $k_F l>2$
described in Sec.~\ref{subsec1} are presented as well. Moreover,
$\tau_\varphi$ behaves naturally with temperature: it increases with
temperature decrease (Fig.\ \ref{fig5}).

Finally, knowing the Drude conductivity we can find the total value of
quantum corrections as $\delta\sigma(T)=\sigma(T)-\sigma_0$. For
$T=0.46$ K, the value of $\delta\sigma$ as a function of $k_F l$ is
depicted in the Fig.\ \ref{fig6}. As the figure illustrates, these
results match those obtained for $k_F l>2$ and discussed in previous
subsection. It is seen that the total quantum correction $\delta\sigma$
is small part of the Drude conductivity $\sigma_0$ at large $k_F l$, is
about 70\% of $\sigma_0$ at $k_F l\simeq 2$, and very close to
$\sigma_0$ at lower $k_F l$. The last leads to the fact that at low
temperature the conductivity in zero magnetic field is very small
fraction of the Drude conductivity (see inset in Fig.~\ref{fig6}). For
instance,  $\sigma/\sigma_0 \simeq 2\times 10^{-3}$ at $k_F l\simeq
0.5$ and $T=0.46$~K, the absolute value of $\sigma$ is about $3\times
10^{-8}$ $\Omega^{-1}$ that is much smaller than $e^2/h\simeq
3.86\times 10^{-5}\, \Omega^{-1}$.

Thus, down to $k_F l \simeq 0.5$ the conductivity looks like diffusive
one and its temperature and magnetic field dependence is due to that of
the quantum corrections which can be comparable in magnitude with the
Drude conductivity at low temperature.

It is usually supposed that at $\sigma<e^2/h$ the conductivity
mechanism is the variable range hopping.\cite{h1,h2,h3,h4,h5} It is
argued by the fact that the temperature dependence of the resistivity
$\rho=\sigma^{-1}$ is well described by characteristic for this
mechanism dependence:
\begin{equation}
\rho(T)=\rho_0 \exp{\left(\frac{T_0}{T}\right)^{m}} \label{eq10}
\end{equation}
with $m=1/2,1/3$ depending on the ratio of the Coulomb gap width to the
temperature. Fig.~\ref{fig8}~(b) shows our experimental results as
$\ln\rho$-versus-$T^{-1/3}$ plot. As seen our data being in an
excellent agreement with Eq.~(\ref{eq09}) [Fig.~\ref{fig8}~(a)] are
well described by Eq.~(\ref{eq10}) too. Surprised by this fact we have
examined some data from Refs.\ \onlinecite{h1,h2,h3,h4,h5}, which were
interpreted from the position of the variable range hopping. We have
found that while the resistivity is less than $\sim 10^6$~$\Omega$
those dependences are well aligned in $[\sigma(T)/G_0 + \ln(
\sigma(T)/G_0)]$-versus-$\ln T$ coordinates also. For example, in
Fig.~\ref{fig8} the data from Ref.~\onlinecite{h5} for carrier density
$1.085\times 10^{11}$ and $0.984\times 10^{11}$ cm$^{-2}$ are
presented. Thus only the temperature dependence of the conductivity
does not allow to identify the conductivity mechanism reliably.

\section{Conclusion}
We have studied the quantum corrections to the conductivity for
gated single quantum well GaAs/InGaAs/GaAs structures with 2D
electron gas. Thorough analysis shows that the temperature, low-
and high-magnetic field dependences of the components of the
conductivity and resistivity tensor are well described within the
framework of the conventional theory of the quantum corrections
down to $k_F l\simeq 2$. At this $k_F l$ the value of the total
correction is not small and is about 70\% of the Drude
conductivity for $T=0.45$ K. It has been shown that for zero
magnetic field the interference contribution to the conductivity
exceeds the contribution due to the electron-electron interaction
in $3-5$ times.

On the further lowering of $k_F l$ down to $\simeq 0.5$ the temperature
and magnetic field dependences of conductivity are in qualitative
agreement with the self-consistent theory by Vollhardt and
W\"{o}lfle,\cite{Voelfle} which is applicable for arbitrary values of
quantum corrections. Thus, in wide range of the low temperature
conductivity starting from $\sigma\simeq 3\times 10^{-8}$ $\Omega^{-1}$
the conductivity is of non-hopping nature. We assume that the
transition from the diffusion to hopping occurs at lower $k_F l$ value.

\subsection*{Acknowledgment}
We are grateful to M.~V.~Sadovskii and I.~V.~Gornyi for useful
discussions and comments. This work was supported in part by the RFBR
through Grants No. 00-02-16215, No. 01-02-06471, and No. 01-02-17003,
the Program {\it University of Russia} through Grants No.~990409 and
No.~990425, the CRDF through Grant No. REC-005, the Russian Program
{\it Physics of Solid State Nanostructures}, and the Russian-Ukrainian
Program {\it Nanophysics and Nanoelectronics}.


\begin{thebibliography}{}
\bibitem{ftn1}When the interference correction is significantly less than
the Drude conductivity the mean free path for calculation of
$B_{tr}$ can be found from $\sigma$ instead of $\sigma_0$.
\end{thebibliography}
\end{document}